\documentclass[prl,twocolumn,superscriptaddress]{revtex4-1}
\usepackage{amssymb}
\usepackage{amsfonts}
\usepackage{bbm}
\usepackage{mathrsfs}
\usepackage{graphics,graphicx,epsfig,bm,amsmath,amsthm,amssymb}
\usepackage{bm}
\usepackage{bbm}
\usepackage{longtable}
\usepackage{multirow}
\usepackage{array}
\usepackage{color}
\usepackage{cases}
\usepackage{booktabs }
\usepackage[usenames,dvipsnames]{xcolor}

\usepackage{xcolor}
\usepackage[none]{hyphenat}
\usepackage{float}
\usepackage{amsmath}
\usepackage[a4paper,colorlinks=true,
linkcolor=blue,citecolor=blue,
pdfauthor={ },
pdftitle={ },
pdfsubject={ },
pdfkeywords={ }]{hyperref}

\begin{document}
\title{Experimental Constraint on an Exotic Parity-Odd Spin- and Velocity-Dependent Interaction with a Single Electron Spin Quantum Sensor}
\author{Man Jiao}
\affiliation{Hefei National Laboratory for Physical Sciences at the Microscale and Department of Modern Physics, University of Science and Technology of China, Hefei 230026, China}
\affiliation{CAS Key Laboratory of Microscale Magnetic Resonance, University of Science and Technology of China, Hefei 230026, China}
\affiliation{Synergetic Innovation Center of Quantum Information and Quantum Physics, University of Science and Technology of China, Hefei 230026, China}

\author{Maosen Guo}
\affiliation{Hefei National Laboratory for Physical Sciences at the Microscale and Department of Modern Physics, University of Science and Technology of China, Hefei 230026, China}
\affiliation{CAS Key Laboratory of Microscale Magnetic Resonance, University of Science and Technology of China, Hefei 230026, China}
\affiliation{Synergetic Innovation Center of Quantum Information and Quantum Physics, University of Science and Technology of China, Hefei 230026, China}

\author{Xing Rong}
\email{xrong@ustc.edu.cn}
\affiliation{Hefei National Laboratory for Physical Sciences at the Microscale and Department of Modern Physics, University of Science and Technology of China, Hefei 230026, China}
\affiliation{CAS Key Laboratory of Microscale Magnetic Resonance, University of Science and Technology of China, Hefei 230026, China}
\affiliation{Synergetic Innovation Center of Quantum Information and Quantum Physics, University of Science and Technology of China, Hefei 230026, China}

\author{Yi-Fu Cai}
\affiliation{CAS Key Laboratory for Research in Galaxies and Cosmology, Department of Astronomy, University of Science and Technology of China, Hefei 230026, China}
\affiliation{School of Astronomy and Space Science, University of Science and Technology of China, Hefei 230026, China}

\author{Jiangfeng Du}
\email{djf@ustc.edu.cn}
\affiliation{Hefei National Laboratory for Physical Sciences at the Microscale and Department of Modern Physics, University of Science and Technology of China, Hefei 230026, China}
\affiliation{CAS Key Laboratory of Microscale Magnetic Resonance, University of Science and Technology of China, Hefei 230026, China}
\affiliation{Synergetic Innovation Center of Quantum Information and Quantum Physics, University of Science and Technology of China, Hefei 230026, China}

\date{\today}

\begin{abstract}
An improved laboratory bound on the exotic spin- and velocity-dependent interaction at micrometer scale is established with a single electron spin quantum sensor.
The single electron spin of a near-surface nitrogen-vacancy center in diamond is utilized as the quantum sensor and a vibrating half-sphere lens is taken as the source of the moving nucleons.
The exotic interaction between the polarized electron and the moving nucleon source is explored by measuring the possible magnetic field felt by the electron spin quantum sensor.
Our experiment set improved constraints on the exotic spin- and velocity-dependent interaction within the force range from 1 to 330 $\mu$m.
The upper limit of the coupling  $g_A^eg_V^N $ at $200 ~\mu m$ is $| g_A^e g_V^N| \leq 8.0\times10^{-19}$, significantly improving the current laboratory limit by
more than four orders of magnitude.
\end{abstract}

\maketitle

Experiments on searching for the interactions mediated by new particles beyond the Standard Model (SM) have received a substantial development in the past decade\cite{safronova2018search,Tanabashi2018}.
Amongst various theoretical models beyond SM, axions are a type of ultra-light and CP-odd scalar fields, which were originally put forward to address the issue of strong CP violation in the SM of particle physics via the Peccei-Quinn mechanism and later introduced as a candidate of dark matter particles\cite{peccei1977cp,weinberg1978new, AxionKim2010}. Accompanied with the novel strategies of detection technology, a wide observational window has been explored that varies from astronomical instruments at cosmological scales to particle colliders at extremely microscopic scales\cite{Bertone2018,Tanabashi2018}. There are many ingenious experiments searching for the axions or axion-like particles mediated force by examining an exotic spin-dependent interaction among electrons and nucleons\cite{Crescini2017, Lee2018,Stadnik2018PRL,Rong2018,Rong2018PRL,Jiao2020}. This type of laboratory experiments were found to be sensitive to the particle physics properties of those hypothetical fields, and hence provides a crucial approach to probing new physics beyond SM\cite{demille2017probing}.

In this paper, we conducted a search for an exotic parity-odd spin- and velocity-dependent interaction between an electron spin and unpolarized nucleon, which is described by an effective potential\cite{dobrescu2006spin}
\begin{equation}
V =  g_A^eg_V^N\frac{\hbar}{4 \pi}(\bm{\sigma} \cdot \bm{v})( \frac{e^{{-\frac{r}{\lambda}}}}{r}) ,
\label{Vexotic}
\end{equation}
where $\bm{\sigma}$ is Pauli vector of the electron spin, $g_A^e$ is the axial-vector coupling constant, $g_V^N$ is the nucleon vector coupling constant, $r=|\textbf{r}|$ with $\textbf{r}$ being the displacement vector between the electron and nucleon,  $\bm{v}$ is their relative velocity, $\lambda=\hbar/(m_\textrm{b}c)$  is the force range with $m_\textrm{b}$ being the boson mass, c being the speed of light in vacuum and $\hbar$ being the Planck's constant divided by $2\pi$. This interaction leads to an effective magnetic field felt by the electron spin arising from the moving nucleons,
\begin{equation}
\textbf{B}(r)=  \frac{g_A^eg_V^N}{2 \pi \gamma_e} \bm{v}  \frac{e^{{-\frac{r}{\lambda}}}}{r},
\label{Beff}
\end{equation}
where $\gamma_e$ is the gyromagnetic ratio of the electron spin.
Recently, experimental upper bounds on this interaction have been set by serval laboratory experiments, such as the atomic parity non-conservation (PNC) experiment \cite{Dzuba2017}, spin-exchange-relaxation-free (SERF) atomic magnetometery one \cite{Kim2019}, and the electron-spin polarized torsion pendulum ones \cite{Heckel2008}.

\begin{figure}
\centering
\includegraphics[width=1\columnwidth]{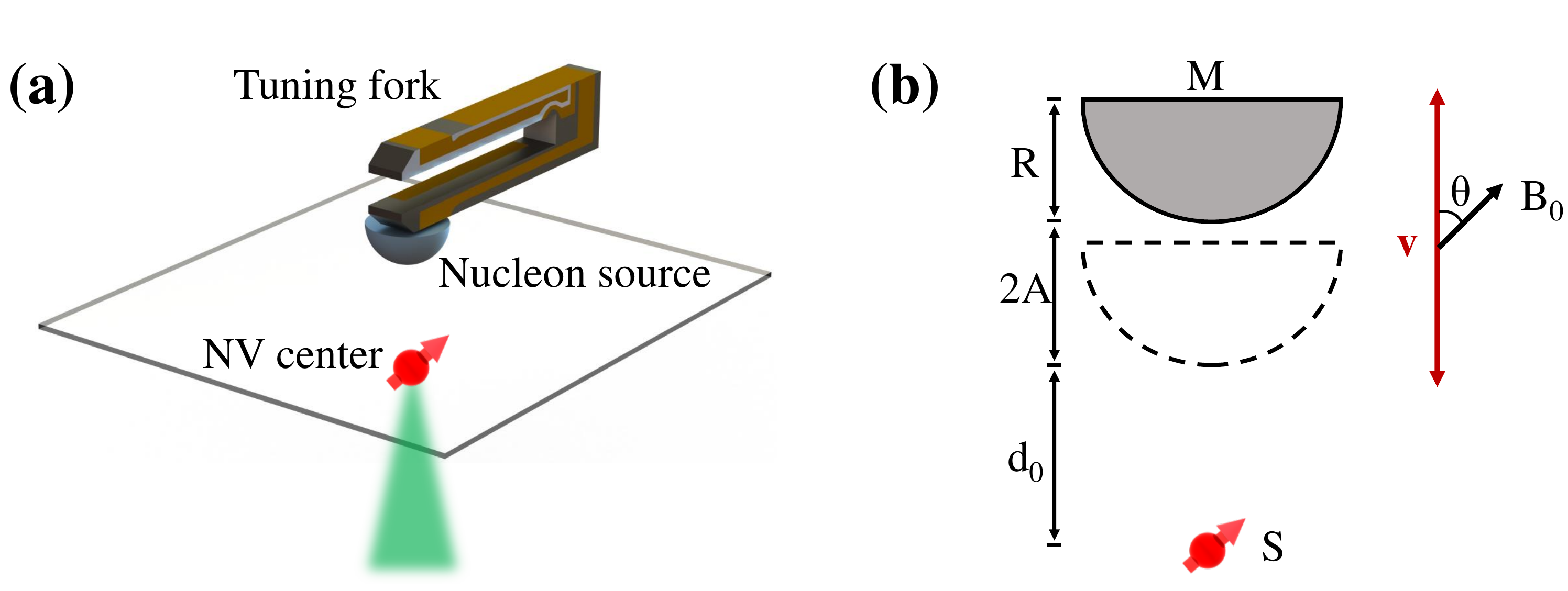}
\caption{(a) Schematic experimental setup. An NV center in diamond is labeled as NV center. A fused silica half-sphere, labeled as Nucleon scource, is utilized as the moving mass source, which is placed on a tuning fork actuator of a AFM. Pulsed green laser has been applied on the NV center for initialization and readout the state of the NV center.
(b) A schematic diagram of interacting source and the single electron spin sensor.
The radius of the nucleon source, M, is $R = 250 ~\mu $m.
A static magnetic field $\mathbf{B_{0}}$ is applied along the symmetry axis of the NV center (S).
$\bm v$ is the relative velocity vector between the NV center and nucleon half-sphere lens.
The angle between the velocity vector and the external magnetic field $\mathbf{B_0}$ is $\theta =\arccos(1/\sqrt{3})$.
}
\label{figure1}
\end{figure}

We utilized a single electron spin of  a near-surface nitrogen-vacancy (NV) center in diamond as a detector for testing the exotic parity-odd interaction. Single NV centers in diamond are defects composed of a substitutional nitrogen atom and a neighboring vacancy\cite{Gruber1997}. They have been applied as nanoscale quantum sensors for detecting weak magnetic field\cite{Schirhagl2014, Degen2017}. Recently, NV centers have been demonstrated as detector for exploring electron-nucleon monopole-dipole interaction\cite{Rong2018} and axial-vector mediated interaction between polarized electrons\cite{Rong2018PRL}. Due to the size of the sensor, which can be engineered to be small compared to the micrometer force range, the geometry of the sensor enables close proximity between the sensor and the source. Furthermore, delicate quantum control method, such as dynamical decoupling techniques\cite{Nature_DDDu}, can be employed to suppress the unwanted magnetic noise, so that the sensitivity of the sensor can be enhanced\cite{Taylor2008}.

Our experiment was carried out on an NV-based magnetometery combined with an atomic force microscope (AFM). The similar setup was utilized for testing the monopole-dipole interaction between an electron spin and nucleons\cite{Rong2018}. Figure 1(a) shows the schematic of our experimental setup. The NV center is close to the surface of the diamond with depth less than 10 nm.  This NV center was created by implantation of 10 keV $N_2^+$ ion into $<$100$>$ bulk diamond and annealing for two hours at $800~^{\circ}$C. After annealing, the diamond was oxidatively etched for 4 hours at $580~^{\circ}$C. We fabricated nanopillars on the surface of the diamond to enhance the detection efficiency of the photoluminescence.  The photoluminescence rate in this experiment has achieved 350 kcounts/s with laser power being about $200~\mu$W. The ground state of the NV center is an electron spin triplet state $^3A_2$ with three substates $|m_S=0\rangle$ and $|m_S=\pm1\rangle$. We applied an external magnetic field along the symmetry axis of the NV center to remove the degeneracy of the $|m_S=\pm1\rangle$ spin states. The strength of the magnetic field is set to be $565~$Gauss, and green laser pulses can initialize the state of NV center to be $|m_S=0\rangle$.
The microwave pulses utilized to manipulate the quantum states of the NV center, are delivered by a copper wire placed on the surface of the diamond. We encode two spin states, $|m_S=0\rangle$ and $|m_S=-1\rangle$, as a quantum sensor, which is sensitive to the magnetic field. The dephasing time of the electron spin obtained from the spin echo\cite{PhysRev.80.580} experiment is $27(4)~\mu$s. The source of the mass is a fused silica half-sphere lens with diameter being $500~\mu$m, which is installed on a tuning fork of the AFM. This is a movable mass source, which enables the detection of the exotic spin- and velocity-dependent interactions. Hereafter, the single electron spin of NV center and the moving mass source are denoted as S and M for convenience, respectively.

\begin{figure*}
\centering
\includegraphics[width=2\columnwidth]{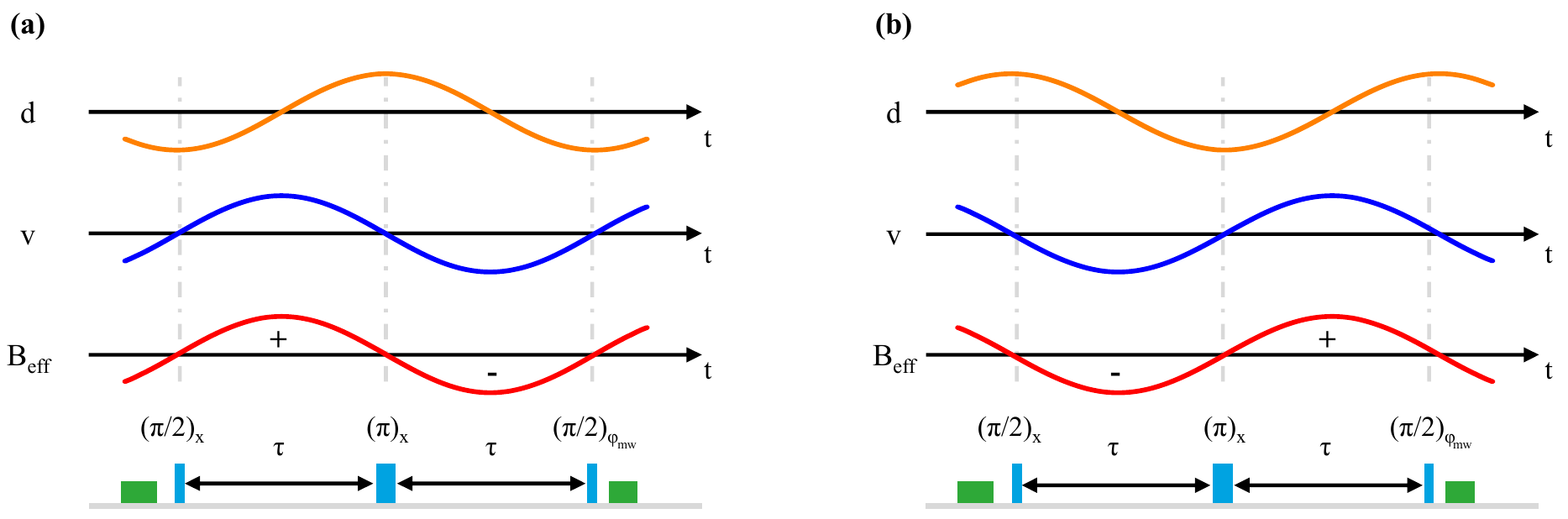}
\caption{Experimental scheme of testing the parity odd spin- and velocity-dependent interaction between a single spin and a moving mass source. (a) and (b) are experimental pulse sequences for accumulating a positive and negative phase factor due to the exotic interaction on the superposition state of S, respectively.  Symbols $\pi$ and $\pi/2$ stand for the rotation angles of the quantum state due to the microwave pulses. The phase of the last $\pi/2$ microwave pulse is $\phi_{mw}$. For accumulating a positive (negative) phase factor, the $\pi/2$ microwave pulses were applied on S when M passed through the minimal (maximum) value of $\text {d(t)}$. $\pi$ microwave pulses were applied on the center of the microwave sequence. The time duration between $\pi/2$ and $\pi$ pulses are $\tau = \pi/\omega_M$.
The pulse lengths of $\pi/2$ and $\pi$ are $64~$ns and $127~$ns , respectively. The time durations of laser for initialization and readout are $2.0~\mu$s and $4.4~\mu$s, respectively. The waiting time $\tau$ is $6.652~\mu$s.
}
\label{figure2}
\end{figure*}

The geometric parameters of the setup are presented in Figure 1(b). Since M is driven by the tuning fork, the distance between S and the bottom of M is $d(t) = d_0 + A[1 + \cos (\omega_M t)]$, where $d_0$ is the minimal distance, A and $\omega_M$ are the amplitude and angular frequency of M, respectively. The velocity of M is $v (t) = A\omega_M\sin(\omega_M t)$. An external magnetic field $\textbf{B}_0$ is applied along the NV axis. The angle between the direction of the velocity, $\bm v$, and the magnetic field, $\textbf{B}_0$, is $\theta =\arccos (1/\sqrt{3})$.
The effective magnetic field on S along the NV axis arising from the hypothetic spin- and velocity-dependent interaction can be derived from integrating $\textbf{B}(r)$ over all nucleons of M as
\begin{equation}
\textbf{B}_{\text{eff}}= \frac{g_A^eg_V^N }{2 \pi \gamma_e}f(\lambda,R,d)\bm{v} \cos\theta,
\end{equation}
where $f(\lambda,R,d)=2 \pi \rho\lambda^2( -e^{-\frac{d+R}{\lambda}}+e^{-\frac{d}{\lambda}}
                                            +\frac{\sqrt{R^2+(d+R)^2}+\lambda}{d+R}e^{-\frac{\sqrt{R^2+(d+R)^2}}{\lambda}}
                                            -\frac{d+\lambda}{d+R}e^{-\frac{d}{\lambda}})$ and $\rho=1.33\times10^{30}~$m$^{-3}$ being the number density of nucleons in M.
(see Supplementary Note 2 for details\cite{SM})

Figure 2 shows the experimental pulse sequence to explore the exotic interaction.  In Figure 2 (a), we show how to accumulate a phase factor due to the possible interaction on the superposition state of S.
The time evolution of distance $\text {d(t)}$ (orange line), the velocity of M, $\text {v(t)}$ (blue line), and the possible $\text B_{\text {eff}}(t)$ (red line) due to the moving nucleons  have been presented in the upper panels, respectively. The laser and the microwave pulse sequences, which were applied on S for initializing, manipulating and readout the states of S, have been shown in the low panel, respectively.  The laser and microwave pulse sequences were synchronized with the vibration of M with a pulse generator and a comparator\cite{Rong2018,Qin2016}.  The first laser pulse and the following $\pi/2$ microwave pulse were utilized to prepare the state of S to a superposition state $(|0\rangle-i|1\rangle)/\sqrt{2}$. The $\pi/2$ microwave pulses were applied on S when M passed through the minimal values of $\text {d(t)}$.  In this case, during the first waiting time $\tau$, the state of spin evolves about the z axis and accumulate a positive phase factor dependent on the magnetic field $\text B_{\text {eff}}$, while S will accumulate a negative phase factor during the second waiting time. Due to the $\pi$ pulse inverting the state of S in the middle of the time evolution, the final state of S will acquire a positive phase factor. After the last $\pi/2$ pulse with a variable phase $\phi_{mw}$ together with a laser pulse, the population of the final state on $|m_S = 0\rangle$ can be written as $P_+ = [1+ \cos(\phi_{mw} + \phi)]/2$ with $\phi =  \int_0 ^\tau \gamma_e \text B_{\text{eff}}(t) dt- \int_{\tau} ^{2\tau} \gamma_e \text B_{\text{eff}}(t) dt$. In figure 2(b), the final state is designed to acquire a negative phase factor with $\pi/2$ pulses being applied on S when M passed through the maximum values of $\text {d(t)}$. The population of the final state on $|m_S = 0\rangle$ in this case can be written as  $P_- = [1+  \cos(\phi_{mw} - \phi)]/2$. In our experiment, we record the difference between the two populations, $I =P_+-P_-= -\sin(\phi_{mw})\sin(\phi)$.

\begin{figure}
\centering
\includegraphics[width=0.95\columnwidth,height=5.5cm]{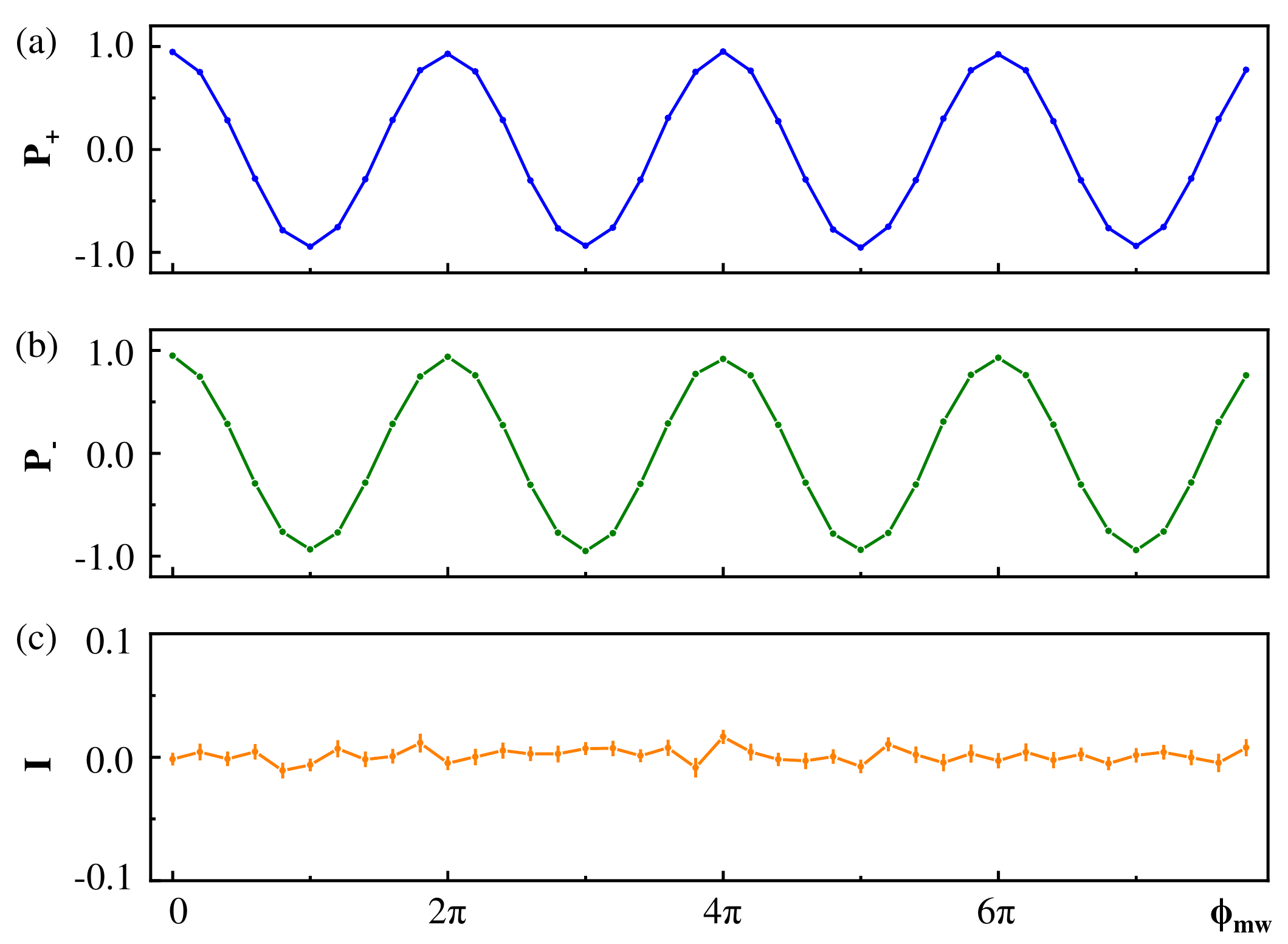}
\caption{ Experimental results for testing the parity-odd spin- and velocity-dependent interaction. (a) and (b) are experimental data for  population of the final state on $|m_S = 0\rangle$,  $P_+ $ and  $P_-$, respectively. (c) Orange dots with error bars are the difference between $P_+$ and  $P_-$.
}
\label{figure3}
\end{figure}

In our experiment, the time duration is $\tau = 6.652~\mu$s, the amplitude of the vibrating M is $A = 165.2(1)~$nm, and the angular frequency $\omega_M = 2\pi \times 74.452~$kHz. The minimal distance between the NV center and the bottom of the half-sphere is $d_0 = 2.0(1)~\mu$m. Figure 3 show the experiment results. We repeated the measurement for 60 millions times to build good statistics. Experiment result shows that the parameter, $\phi = 0.0011\pm 0.0014$, can be obtained by fitting the data with $I = -\sin(\phi_{mw})\sin(\phi)$. There is no exotic parity-odd spin- and velocity-dependent interaction observed at current experiment. Nevertheless, an experimental limit on such interaction can be obtained from our experiment.
\begin{table}
 \caption{Summary of the systematic errors in our experiments. The corrections to $g_\textrm{A}^\textrm{e}g_\textrm{V}^\textrm{N}$ with $\lambda = 200~\mu$m are listed.}
 \label{table1}
 \begin{tabular}{l c c}
  \hline
  \hline
 Systematic error  &Size of effect & Corrections \\
  \hline
The angle $\theta $ & $54.7\pm0.6^{\circ}$ & $(0.2\pm9.7)\times10^{-21}$ \\
Distance between M and S & $2.0\pm0.1~\mu$m & $(0.0\pm 6.3)\times 10^{-22}$ \\
Diameter of M& $500\pm 2.5~\mu$m  & $(0.02 \pm 4.14)\times 10^{-21} $ \\
Thickness of M& $250\pm 35~\mu$m  & $(1.0\pm5.3)\times10^{-20}$ \\
Amplitude of vibration&$165.2\pm0.1~nm$&$(0.0\pm4.1)\times10^{-22}$\\
Deviation in x-y plane&$1.3\pm0.8~\mu m$&$(3.1\pm3.1)\times10^{-23}$\\  \hline
\hline
  Total&&$(1.0\pm5.4)\times10^{-20}$\\
  \hline
  \hline
\end{tabular}
\end{table}

Table 1 provides the systematic error budget of our experiment, where we take $\lambda = 200~\mu$m as an example. The systematic error due to the uncertainty of the angle between $\textbf{B}_\text{eff}$ and the NV axis  leads to the correction to the coupling being $(0.2\pm9.7)\times 10^{-21}$. The minimal distance between M and S is $(2.0\pm0.1)~\mu$m. The correction to the coupling due to the uncertainty of the distance is estimated to be $(0.0\pm6.3)\times 10^{-22}$.  The amplitude of the vibration of M is $165.2\pm0.1~$nm, and the  correction to the coupling is $(0.0\pm4.1)\times10^{-22}$. The uncertainty of the diameter of the half-sphere lens leads to the correction to the coupling being $(0.02\pm4.14)\times 10^{-21}$. The uncertainty of thickness of the nucleon source has also been taken into account. Since our experiment has been conducted in an external magnetic field being $ 565~$Gauss, one possible systematic error is due to the diamagnetism of M and the tuning fork.  In Supplemental material, we show that the effect of the modulated magnetic field due to the diamagnetism of M and the tuning fork can be canceled by our experimental sequence\cite{SM}.  Since the NV center didn't locate exactly under the center of the half-sphere, the misalignment between the NV center  and the half-sphere in x-y plane is also considered as a source of systematic error.  The total correction to $g_A^eg_V^N$ due to the systematic errors is $(1.0\pm5.4)\times10^{-20}$. The bound from our experiment for the interaction with the force range being $200~\mu$m, is  $|g_A^eg_V^N|\leq8.0\times10^{-19}$ with a $95\%$ confidence level, when both statistical and systematic errors are taken into account. The other values of upper bound with different values of force range can be obtained with the same method.

\begin{figure}
\centering
\includegraphics[width=0.95\columnwidth]{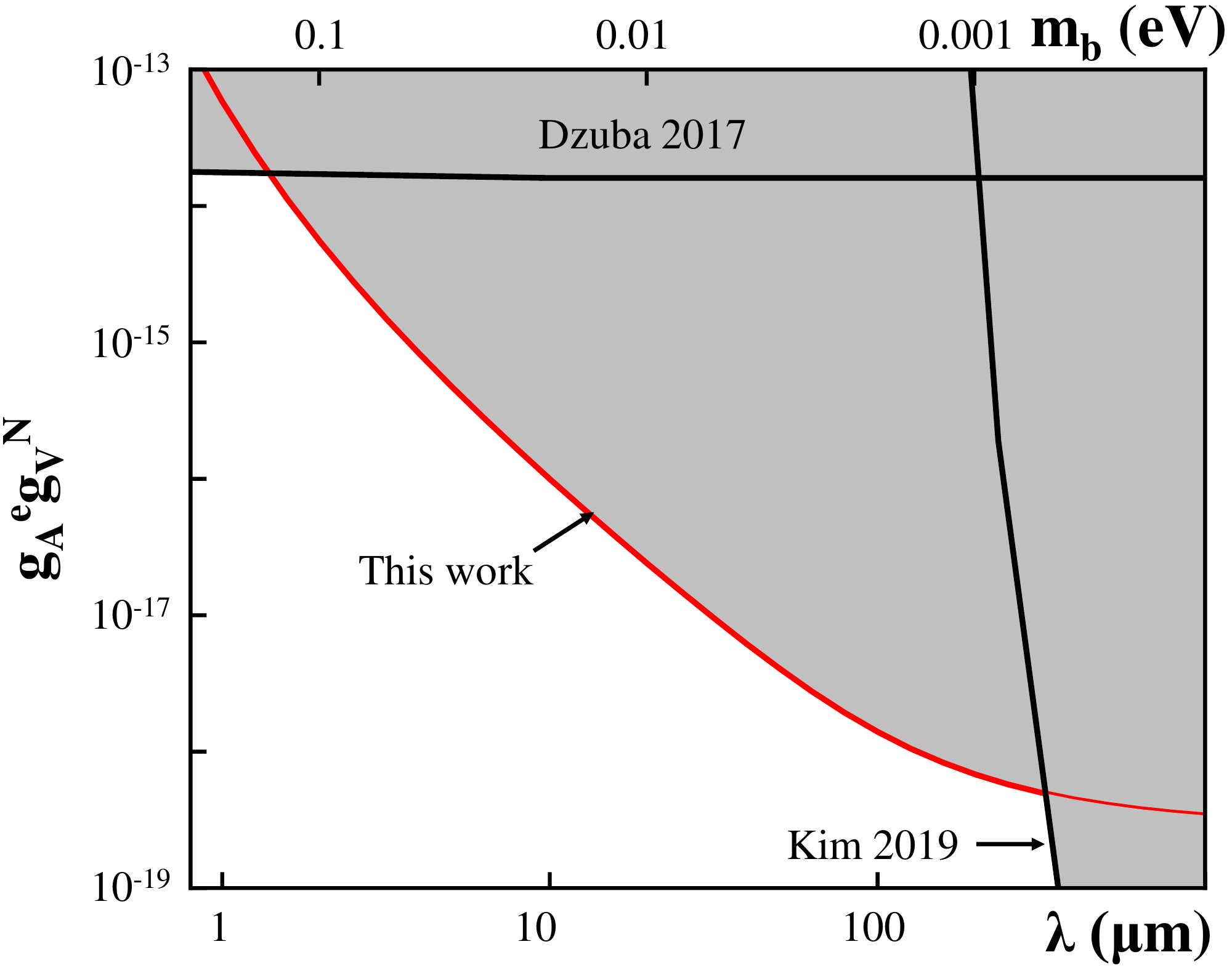}
\caption{Upper limit on the exotic parity-odd spin- and velocity-dependent interaction,  $g_A^eg_V^N $, as a function of the force range $\lambda$ and  mass of the bosons $m_\textrm{b}$. Black lines are upper limits established by experiments in Refs.\cite{Dzuba2017,Kim2019}. The red line is the upper bound obtained from our experiment, which establishes an improved laboratory bound in the force range from 1 to 330 $\mu$m. The upper limit of the coupling  $g_A^eg_V^N $ at $200 ~\mu m$ is $|g_A^eg_V^N | \leq 8.0\times10^{-19}$, which is significantly improved by up to four orders of magnitude.}
\label{figure4}
\end{figure}

Figure 4 shows the upper bound on the parity-odd spin- and velocity-dependent interaction between polarized electron and unpolarized nucleon established by this work together with recent constraints. Grey filled areas are excluded values from experimental searches.  For force range $\lambda > 330~\mu$m, the strongest constraint was set by Kim \emph{et al}.  in Ref.\cite{Kim2019}, and for the force range $\lambda < 1~\mu$m, the upper bound was established by Dzuba \emph{et al}. in Ref.\cite{Dzuba2017}. For the force range from 1 to 330$~\mu$m, the best experimental bound is obtained from our experiment as the red line shown in the Figure 4. The experimental upper limit for the force range $\lambda = 200~\mu$m,  is $|g_A^eg_V^N| \leq 8.0   \times10^{-19}$, which is more than four orders of magnitude more stringent than the bound established  in the previous result in Ref. \cite{Dzuba2017}.

In summary, we report an experimental search for a type of parity-odd spin- and velocity-dependent interaction. A single electron spin of NV center in diamond has been utilized as a sensitive quantum sensor for detecting the possible magnetic field due to the exotic interaction between the electron spin and a moving mass source. The current experimental sensitivity is mainly limited by the sensitivity of the NV center quantum sensor, which  can be further improved by enhance the coherence time of the NV center. Our experimental setup can be further developed to search for exotic parity-even spin- and velocity-dependent interactions between polarized electrons and nucleons.  Our result shows that NV based quantum sensing setup can be utilized as a promising platform not only for physics within the standard model but also for searching  interactions predicted by physics beyond the standard model.

This work was supported by the National Key R$\&$D Program of China (Grant Nos. 2018YFA0306600 and No. 2016YFB0501603), the National Natural Science Foundation of China (Grants Nos. 11722327, 11961131007 and 11653002), the Chinese Academy of Sciences (Grants No. GJJSTD20170001, No. QYZDY-SSW-SLH004 and No. QYZDB-SSW-SLH005), and Anhui Initiative in Quantum Information Technologies (Grant No. AHY050000).
X.\ R thank the Youth Innovation Promotion Association of Chinese Academy of Sciences for the support.
Y. F. C. is supported in part by the CAST Young Elite Scientists Sponsorship Program (2016QNRC001) and by the Fundamental Research Funds for the Central Universities.
Man Jiao and Maosen Guo contributed equally to this work.



\begin{thebibliography}{10}
\expandafter\ifx\csname url\endcsname\relax
  \def\url#1{\texttt{#1}}\fi
\expandafter\ifx\csname urlprefix\endcsname\relax\def\urlprefix{URL }\fi
\providecommand{\bibinfo}[2]{#2}
\providecommand{\eprint}[2][]{\url{#2}}



\bibitem{safronova2018search}
M. S. Safronova, D. Budker, D. DeMille, D. F. Jackson Kimball, A. Derevianko, and C.W. Clark,
\newblock \emph{\bibinfo{journal}{Rev. Mod. Phys.}}
  \textbf{\bibinfo{volume}{90}}, \bibinfo{pages}{025008}
  (\bibinfo{year}{2018}).

\bibitem{Tanabashi2018}
\bibinfo{author}{M.Tanabashi} \emph{et~al.},
\newblock \emph{\bibinfo{journal}{Phys. Rev. D}}
  \textbf{\bibinfo{volume}{98}}, \bibinfo{pages}{030001}
  (\bibinfo{year}{2018}).

\bibitem{peccei1977cp}
\bibinfo{author}{R. D. Peccei} and \bibinfo{author}{H. R. Quinn},
\newblock \emph{\bibinfo{journal}{Phys. Rev. Lett.}}
  \textbf{\bibinfo{volume}{38}}, \bibinfo{pages}{1440}
  (\bibinfo{year}{1977}).

\bibitem{weinberg1978new}
\bibinfo{author}{S.Weinberg},
\newblock \emph{\bibinfo{journal}{Phys. Rev. Lett.}}
  \textbf{\bibinfo{volume}{40}}, \bibinfo{pages}{223} (\bibinfo{year}{1978}).

\bibitem{AxionKim2010}
\bibinfo{author}{J. E. Kim} and \bibinfo{author}{G. Carosi},
\newblock \emph{\bibinfo{journal}{Rev. Mod. Phys.}}
  \textbf{\bibinfo{volume}{82}}, \bibinfo{pages}{557}
  (\bibinfo{year}{2010}).


\bibitem{Bertone2018}
\bibinfo{author}{G. Bertone} and  \bibinfo{author}{M. P. Tim},
\newblock \emph{\bibinfo{journal}{ Nature} }
  \textbf{\bibinfo{volume}{562}}, \bibinfo{pages}{51}
  (\bibinfo{year}{2018}).



\bibitem{Crescini2017}
N. Crescini,C. Braggioa, G. Carugnoa, P. Falferib, A. Ortolanc, and G. Ruoso,
\newblock\emph{\bibinfo{journal}{Nucl. Instrum. Methods Phys Res. Sec. A}}
\textbf{\bibinfo{volume}{842}}, \bibinfo{number}{109}
(\bibinfo{year}{2017}).


\bibitem{Lee2018}
\bibinfo{author}{J. Lee}, \bibinfo{author}{A. Almasi} and \bibinfo{author}{M.Romalis},
\newblock\emph{\bibinfo{journal}{Phys. Rev. Lett.}}
\textbf{\bibinfo{volume}{120}}, \bibinfo{number}{161801}
(\bibinfo{year}{2018}).


\bibitem{Stadnik2018PRL}
\bibinfo{author}{Y. V. Stadnik},\bibinfo{author}{V. A. Dzuba}, and \bibinfo{author}{V. V. Flambaum},
\newblock\emph{\bibinfo{journal}{Phys. Rev. Lett.}}
\textbf{\bibinfo{volume}{120}}, \bibinfo{number}{013202}
(\bibinfo{year}{2018}).



\bibitem{Rong2018}
\bibinfo{author}{X. Rong} \emph{et~al.},
\newblock\emph{\bibinfo{journal}{Nat. Commun.}}
\textbf{\bibinfo{volume}{9}}, \bibinfo{number}{739}
(\bibinfo{year}{2018}).

\bibitem{Rong2018PRL}
X. Rong, M. Jiao, J. Geng, B. Zhang, T. Xie, F. Shi, C.-K. Duan, Y.-F. Cai, and J. Du,
   \newblock\emph{\bibinfo{journal}{Phys. Rev. Lett.}}
   \textbf{\bibinfo{volume}{121}}, \bibinfo{number}{080402}
(\bibinfo{year}{2018}).

\bibitem{Jiao2020}
M. Jiao, X. Rong, H. Liang,Y. Cai, and J. Du,
   \newblock\emph{\bibinfo{journal}{Phys. Rev. D}}
   \textbf{\bibinfo{volume}{101}}, \bibinfo{number}{115011}
(\bibinfo{year}{2020}).

\bibitem{demille2017probing}
\bibinfo{author}{D. DeMille}, \bibinfo{author}{J.M. Doyle}, and \bibinfo{author}{A.O. Sushkov},
\newblock \emph{\bibinfo{journal}{ Science} }
  \textbf{\bibinfo{volume}{357}}, \bibinfo{pages}{990}
  (\bibinfo{year}{2017}).

\bibitem{dobrescu2006spin}
B. A. Dobrescu and I. Mocioiu,
\newblock \emph{\bibinfo{journal}{Journal of High Energy Physics}}
  \textbf{\bibinfo{volume}{2006}}, \bibinfo{pages}{005} (\bibinfo{year}{2006}).


\bibitem{Dzuba2017}
\bibinfo{author}{V.A. Dzuba},
\bibinfo{author}{V.V. Flambaum}, and \bibinfo{author}{Y.V. Stadnik},
\newblock\emph{\bibinfo{journal}{Phys. Rev. Lett.}}
\textbf{\bibinfo{volume} {119}}, \bibinfo{number}{223201}
(\bibinfo{year} {2017}).

\bibitem{Kim2019}
\bibinfo{author}{Y. Kim},
\bibinfo{author}{P. Chu},
\bibinfo{author}{I. Savukov},
and \bibinfo{author}{S. Newman},
\newblock \emph{\bibinfo{journal}{Nature Communications}}
\textbf{\bibinfo{volume}{10}}, \bibinfo{pages}{2245}
(\bibinfo{year}{2019}).


\bibitem{Heckel2008}
\bibinfo{author}{B. R. Heckel},\bibinfo{author}{E. G. Adelberger}, and \bibinfo{author}{C. E. Cramer},
 \newblock\emph{\bibinfo{journal}{Phys. Rev. D}}
 \textbf{\bibinfo{volume}{78}}, \bibinfo{number}{092006}
 (\bibinfo{year}{2008}).

\bibitem{Gruber1997}
\bibinfo{author}{A. Gruber}, \bibinfo{author}{A. Dräbenstedt}, \bibinfo{author}{C. Tietz}, \bibinfo{author}{L. Fleury, J. Wrachtrup} and \bibinfo{author}{C. von Borczyskowski},
\newblock\emph{\bibinfo{journal}{Science}}
 \textbf{\bibinfo{volume}{276}}, \bibinfo{number}{2012}
 (\bibinfo{year}{1997}).

\bibitem{Schirhagl2014}
\bibinfo{author}{R. Schirhagl},\bibinfo{author}{K. Chang},\bibinfo{author}{M. Loretz},\bibinfo{author}{C. L. Degen},
 \newblock\emph{\bibinfo{journal}{Annu. Rev. Phys. Chem.}}
 \textbf{\bibinfo{volume}{65}}, \bibinfo{number}{83}
 (\bibinfo{year}{2014}).


\bibitem{Degen2017}
\bibinfo{author}{C. L. Degen},\bibinfo{author}{F. Reinhard},and \bibinfo{author}{P. Cappellaro},
 \newblock\emph{\bibinfo{journal}{Rev. Mod. Phys.}}
 \textbf{\bibinfo{volume}{89}}, \bibinfo{number}{1}
 (\bibinfo{year}{2017}).


\bibitem{Nature_DDDu}
J. Du, X. Rong, N. Zhao, Y. Wang, J. Yang, and R. B. Liu,
\newblock \emph{\bibinfo{journal} {Nature(London)}}
\textbf{\bibinfo{volume}{461}}, \bibinfo{number}{1265}
(\bibinfo{year}{2009}).

\bibitem{Taylor2008}
J. M. Taylor, P. Cappellaro, L. Childress, L. Jiang, D. Budker, P. R. Hemmer, A. Yacoby, R. Walsworth, and M. D. Lukin,
\newblock \emph{\bibinfo{journal}{Nat. Phys.}}
\textbf{\bibinfo{volume}{4}}, \bibinfo{number}{810}
(\bibinfo{year}{2008}).


\bibitem{PhysRev.80.580}
\bibinfo{author}{E.Hahn},
\newblock \emph{\bibinfo{journal}{Phys. Rev.}}
\textbf{\bibinfo{volume}{80}}, \bibinfo{number}{580}
(\bibinfo{year}{1950}).


\bibitem{Qin2016}
X. Qin, Y. J. Xie, R. Li, X. Rong, X. Kong, F. Z. Shi, P. F. Wang, and J. F. Du,
\newblock \emph{\bibinfo{journal}{IEEE Magn. Lett.}}
\textbf{\bibinfo{volume}{7}}, \bibinfo{number}{1}
(\bibinfo{year}{2016}).

\bibitem{SM}
\newblock \bibinfo{title}{Supplemental materials for details of the experimental setup and the statistical and systematic error analysis.}


\end{thebibliography}
\end{document}